\documentclass[final,5p,times,twocolumn]{elsarticle}
\usepackage{graphicx}
\usepackage{amssymb}
\usepackage{amsmath}
\usepackage{bm}
\usepackage{multirow,bigdelim}
\usepackage{xfrac}

\usepackage{booktabs}
\usepackage{array}
\newcolumntype{L}{>{\centering\arraybackslash}m{8cm}}

\biboptions{compress}

\usepackage{pdflscape}

\usepackage[breaklinks=true]{hyperref}
\hypersetup{pdftitle={ElecSus2}
}

\newcounter{bla}

\newcommand{\mrm}{\mathrm}

\journal{Computer Physics Communications}

\begin{document}

\begin{frontmatter}



\title{ElecSus: Extension to arbitrary geometry magneto-optics}


\author{James Keaveney \corref{author}}
\author{Charles S. Adams}
\author{Ifan G. Hughes}

\cortext[author]{Corresponding author.\textit{E-mail address:} james.keaveney@durham.ac.uk}
\address{Joint Quantum Centre (JQC) Durham-Newcastle, Department of Physics, Durham University, South Road, Durham, DH1 3LE, United Kingdom}

\begin{abstract}
We present a major update to ElecSus, a computer program and underlying model to calculate the electric susceptibility of an alkali-metal atomic vapour. Knowledge of the electric susceptibility of a medium is essential to predict its absorptive and dispersive properties.
In this version we implement several changes which significantly extend the range of applications of ElecSus, the most important of which is support for non-axial magnetic fields (i.e. fields which are not aligned with the light propagation axis). Suporting this change requires a much more general approach to light propagation in the system, which we have now implemented. We exemplify many of these new applications by comparing ElecSus to experimental data.
In addition, we have developed a graphical user interface front-end which makes the program much more accessible, and have improved on several other minor areas of the program structure.

%
%
%
\end{abstract}

\begin{keyword}
Spectroscopy \sep Faraday effect \sep Atom-light interaction \sep Alkali atom \sep FADOF \sep Magneto-Optics \sep Voigt effect \sep Electric field propagation \sep Polarimetry \sep Stokes Parameters 

\end{keyword}

\end{frontmatter}



{\bf PROGRAM SUMMARY}

\begin{small}
\noindent
{\em Program Title:} ElecSus                               \\
{\em Licensing provisions:} Apache License, Version 2.0       \\
{\em Programming language:} Python                                \\
{\em Computer:} Any single computer running Python 2.           \\
{\em Operating system:} Linux, Mac OSX, Windows.                       \\
{\em RAM:} Depends on the precision required and size of the data set, but typically not larger than 200 MiB with GUI, 50 MiB as a function call.     \\
{\em Number of processor cores used:} From 1 to all available, depending on the fitting method. Single lineshape calculations use a single core. \\
{\em Keywords:} Spectroscopy, Faraday effect, Atom-light interaction, Alkali atom, FADOF, Magneto-Optics, Voigt Effect, Electric field propagation, Polarimetry, Stokes Parameters  \\
{\em Classification:} 2.2, 2.3                                       \\
{\em External routines/libraries:} SciPy library~[1] 0.15.0 or later, NumPy~[1], matplotlib~[2], sympy~[3], lmfit 0.9.5 or later~[4], wxpython (required for GUI only)    \\
{\em Nature of problem:}\\
Calculating the weak-probe electric susceptibility of an alkali-metal vapour. The electric susceptibility can be used to calculate spectra such as transmission and Stokes parameters. Measurements of experimental parameters can be made by fitting the theory to data.
   \\
{\em Solution method:}\\
  The transition frequencies and wavelengths are calculated using a matrix representation of the Hamiltonian in the completely uncoupled basis. A suite of fitting methods are provided in order to allow user supplied experimental data to be fit to the theory, thereby allowing experimental parameters to be extracted.
  \\
{\em Restrictions:}\\
Results are only valid in the weak-probe regime.
   \\
{\em Running time:}\\
Depends on the number of data points, but typically less than a second for a theory curve in the Faraday or Voigt geometry. Other geometries will take longer. Fitting will take anywhere from 10 seconds to 20 minutes depending on the method used, the number of parameters to fit and the number of data points.
   \\

\end{small}


\section{Introduction}
\label{sec:Introduction}


The fundamental interaction between atoms and light continues to underpin a great deal of scientific research. In atomic vapours, understanding and control over this interaction has enabled a vast array of applications, including  
compact atomic clocks~\cite{Knappe2004}, 
magnetometers~\cite{Schwindt2007,Knappe2016}, magnetoencephalography~\cite{Sander2012,Alem2015}, thermometry~\cite{Truong2011}, laser frequency stabilisation both on~\cite{Affolderbach2005} and off-resonance~\cite{Zentile2014a,Keaveney2016c}, enhanced frequency up-conversion~\cite{Vernier2010}, trans-spectral orbital angular momentum transfer~\cite{Walker2012} and quantum memories~\cite{Heshami2015}.

Development of computational tools such as ARC~\cite{Sibalic2016a}, The Software Atom~\cite{Javanainen2016a} and ElecSus~\cite{Zentile2015b} 
plays an important role in the development of these applications, where system parameters can be optimised in theory then tested and verified experimentally. 


For example, understanding the absorption and dispersion of an atomic vapour has led to a deeper understanding of atomic filters based on the Faraday effect, 
and modelling this has led to the optimisation of the linewidth of such filters~\cite{Zentile2015c}, 
or optimisation of the filter in the presence of homogeneous broadening~\cite{Zentile2015d}. 
One can then use these optimised filters in other applications, e.g. using the filter to make an intrinsically frequency-stable laser system~\cite{Keaveney2016c}, creating a dichroic beam splitter for Raman light~\cite{Abel2009a} or filtering frequency-degenerate photon pairs from an optical parametric oscillator~\cite{Zielinska2014}.

The previous version of ElecSus has been used in a wide range of experiments, including magnetic field imaging~\cite{Horsley2015,Horsley2016}, Faraday filtering~\cite{Zentile2015c,Zentile2015d,Keaveney2016c,Portalupi2016,Kiefer2016}, characterisation of hybrid atom-cavity systems~\cite{Munns2016}, determination of spin polarization of optically pumped atoms~\cite{Ding2016} and absolute absorption measurements~\cite{Hanley2015}.
Since the first publication of ElecSus in 2015~\cite{Zentile2015b}, we have added significant functionality that adds to both the scientific scope and the accessibility of the program. In brief, these are:
\begin{itemize}
\item Adding support for magnetic fields that are not parallel to the light propagation axis (i.e. non-Faraday geometry). 
\item Directly calculating the propagation of electric fields via Jones calculus, which allows, for example: magnetic field gradients across the atomic medium; simulating imperfect polarisers; and simulating birefringent optics.
\item A graphical user interface (GUI) now allows users with no knowledge of computer programming to use the majority of the program features.
\item A rewrite of the fitting methods using the {\it lmfit} module~\cite{Newville2014}, which allows the user to impose bounds on each of the fit parameters. In addition, the user can now select the {\tt differential\_evolution} fitting algorithm which is an efficient global optimisation routine for multi-parameter fits.
\item Several  minor changes and bug fixes to program operation since original publication
\end{itemize}
In the remainder of the paper, we discuss the physics and computational implementation of the above additions.

\section{Recap of important concepts}
\label{sec:Theory}

The majority of the theoretical background is unchanged from the original ElecSus publication~\cite{Zentile2015b}. However, we briefly summarise here the general principles and important equations.

The main effort of the program is to calculate the linear electric susceptibility, $\chi$, of an atomic vapour exposed to a near-resonant weak-probe laser field. In the weak-probe limit~\cite{Smith2004,Sherlock2009} optical pumping can be neglected and the optical properties of nearby transitions can be treated independently. A single transition,  labelled $i$, is treated as an isolated two-level atom, where neglecting the atomic motion, the susceptibility is given by
\begin{align}
\chi_i(\Delta_i) &= \frac{C_i^2d^2\mathcal{N}_g}{\varepsilon_0\hbar}f(\Delta_i), \\
f(\Delta_i) &= \frac{\mrm{i}}{\Gamma/2-\mrm{i}\Delta_i},
\label{eq:Chi_bare}
\end{align}
where $d^2$ is the reduced dipole matrix element of the transition, $C_i^2$ is the transition strength, $\Gamma$ is the natural linewidth of the transition, $\Delta_i = \omega - \omega_i$ is the difference between the laser angular frequency $\omega$ and the resonance frequency $\omega_i = (E_e - E_g)/\hbar$, which is the difference in energy between a ground state $\vert g \rangle$ and excited state $\vert e \rangle$ divided by the reduced Planck constant.  
$\mathcal{N}_{g}$ is the atomic number density of a particular state in the ground manifold, which in thermal equilibrium is given by
\begin{align}
\mathcal{N}_g = F_a \mathcal{N} \frac{\exp(-\Delta E_g / k_{\rm B} T)}{\sum_{j=1}^{2(2I+1)}  \exp(-\Delta E_j / k_{\rm B} T)},
\end{align}
where $\mathcal{N}$ is the total atomic number density, $F_a$ is the isotopic fraction, $k_{\rm B}$ is Boltzmann's constant, and $T$ is the absolute temperature in Kelvin. The sum is over all of the nuclear sub-states, labelled according to the nuclear spin quantum number $I$ and its projection $m_I$. The energy difference $\Delta E_j$ is measured with respect to the lowest energy in the ground manifold. The fractional weighting of each of the $2(2I+1)$ ground states is nearly uniform, i.e. $1/(2(2I+1))$, for most cases, except where the temperature is very low or the energy difference between ground states becomes very large, for example in  extremely high ($\gtrsim 1$~T) magnetic fields. This fractional weighting is a new addition since the original version of ElecSus~\cite{Zentile2015b}.

Atomic motion causes an inhomogeneous broadening of the bare atomic lines; the atoms experience a Doppler-shifted frequency according to their component of velocity in the direction of the beam $v_z$, which is a 1D Maxwell-Boltzmann distribution given by equation (6) in the original publication~\cite{Zentile2015b}. The resulting atomic lineshape is thus a convolution between the stationary atomic response (the Lorentzian $f(\Delta)$) and the Gaussian distribution of velocities $g(v)$, which yields the well-known Voigt profile,
\begin{align}
\chi_i(\Delta_i) &= \frac{C_i^2d^2\mathcal{N}_a}{\epsilon_0\hbar}\mathcal{V}(\Delta_i), \label{eq:Chi_Doppler} \\
\mathcal{V}(\Delta_i) &= \int_{-\infty}^\infty f(\Delta_i-kv)g(v)\mrm{d}v.
\end{align}
In a multi-level system, the total susceptibility at a given global frequency detuning $\Delta$ can be found by summing over each transition,
\begin{align}
\chi(\Delta) = \sum_i \chi_i(\Delta - \Delta_i).
\end{align}
We use a matrix representation of the atomic Hamiltonian in the $\vert m_L, m_S, m_I \rangle$ quantum number basis ($m_{L,S,I}$ are the quantum numbers associated with the projection of the electronic orbital, electronic spin and nuclear spin angular momenta, respectively) to calculate the resonant frequencies and transition strengths, as described in section 2.2 of the original publication~\cite{Zentile2015b}. The Hamiltonian includes details of the internal level structure such as fine and hyperfine structure, and interactions with an external magnetic field (Zeeman effect). A separate Hamiltonian is calculated for the ground state $n$S and excited state $n$P; the Hamiltonians are diagonalised to find eigenenergies $E_j$ and eigenstates $\vert j \rangle$, which are in general a superposition of basis states. The transition frequencies are the difference in energy between a ground state $\vert g \rangle$ and excited state $\vert e \rangle$. 
The transition strength is calculated from the dipole matrix element $\vert \langle g\vert {\rm e}r_q \vert e \rangle \vert^2$, where the subscript $q$ denotes the component of the dipole operator in the spherical basis, and denotes the type of electronic transition; $\pi$, $\sigma^+$ or $\sigma^-$ which are associated with a $\Delta m_L = 0, +1$ or $-1$ transition, respectively. 
In the previous version of ElecSus~\cite{Zentile2015b} the $\pi$ transitions were not calculated since, in the Faraday geometry where the magnetic field vector is parallel to the wavevector of the light ($\hat{B} \cdot \hat{k} = \pm 1$, where the hat denotes the unit vector, $\hat{B}=\vec{B}/\vert B \vert$), $\pi$ transitions are forbidden~\cite{Bransden2003}. 
In this version we relax the constraint on the magnetic field geometry, and therefore we additionally calculate the $\pi$ transitions.

\section{Electric field propagation in an atomic medium with an applied magnetic field}

The major improvement over the original version of ElecSus (versions 1 and 2) is relaxing the constraint that the magnetic field axis must be parallel to the light propagation axis (i.e. the Faraday geometry). 
In making this change, we must also consider how the polarisation of the input light couples to the atomic transitions, and the resulting effect this has on light propagation through the atomic medium. In this section we will present the general approach to this problem, and then the special cases of the Faraday and Voigt geometries. This section follows the work of Palik and Furdyna~\cite{Palik1970}, and Rotondaro, Zhdanov and Knize~\cite{Rotondaro2015}.

\subsection{The wave equation and the dielectric tensor}

We start with Maxwell's wave equation for a non-magnetic dielectric medium, which is given by
\begin{align}
\vec{k} \times (\vec{k} \times \vec{E}) + \frac{\omega^2}{\epsilon_0 c^2} \epsilon \cdot \vec{E} = 0,
\end{align}
where $\vec{k}$ is the wavevector, $\vec{E}$ is the electric field, $\omega$ is the angular frequency of the plane-wave and $\epsilon$ is the dielectric tensor.

\begin{figure}
	\begin{center}
	\includegraphics[width=0.65\columnwidth]{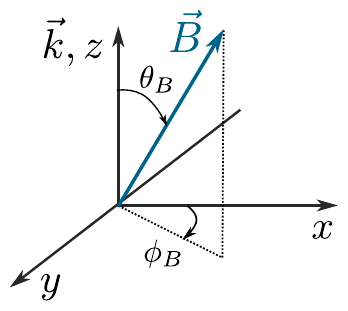}
	\end{center}
	\caption{Representation of the geometry of the system under consideration. $\theta_B$ is the angle that $\vec{B}$ makes with the $z$-axis, $\phi_B$ is the angle between $x$ and the projection of $\vec{B}$ into the $x-y$ plane, and the arrow-heads indicate the sign convention. $\theta_B = 0, \phi_B = 0$ yields the Faraday geometry, while $\theta_B=\pi/2$ or $\phi_B = \pi/2$ becomes the Voigt geometry.}
	\label{fig:geometry}
\end{figure}

We assume that the light is a plane-wave that propagates in the $z$-axis; the electric field therefore lies in the $x-y$ plane. The applied magnetic field $\vec{B}$ can take any angle, but it is practically easiest to rotate the coordinate system around the $z$-axis such that the magnetic field lies in the $x-z$ plane, which is effectively just a polarisation rotation in the $x-y$ plane by an angle $\phi_B$. The magnetic field then makes an angle $\theta_B$ with the $z$-axis, as shown in figure~\ref{fig:geometry}.

After defining the complex refractive index of the medium $n$,
\begin{align}
n^2 = \left( \frac{c}{\omega} \right)^2 \vec{k} \cdot \vec{k},
\end{align}
in the coordinate system discussed above, the wave equation can be written in matrix form as~\cite{Rotondaro2015}
\begin{align}
\left(
\begin{array}{ccc}
( \epsilon_x - n^2) \cos(\theta_B) & \epsilon_{xy} & \epsilon_x \sin(\theta_B) \\
-\epsilon_{xy} \cos(\theta_B)  & \epsilon_x - n^2 & -\epsilon_{xy} \sin(\theta_B) \\
(n^2 - \epsilon_z) \sin(\theta_B) & 0 & \epsilon_z \cos(\theta_B)
\end{array}
\right)
\left(
\begin{array}{ccc}
E_x \\ E_y \\ E_z
\end{array}
\right)
= 0,
\label{eq:wavematrix}
\end{align}
where $\epsilon_x$, $\epsilon_{xy}$, and $\epsilon_z$ are elements of the dielectric tensor, and are related to the complex electric susceptibility by
\begin{subequations}
\begin{align}
\epsilon_x &= \frac{1}{2} ( 2 + \chi_+ + \chi_- ), \\
\epsilon_{xy} &= \frac{\rm i}{2} (\chi_- - \chi_+ ), \\
\epsilon_z &= 1 + \chi_0,
\end{align}
\end{subequations}
and the $\chi_{+,-,0}$ are the susceptibilities associated with $\sigma^+$, $\sigma^-$ and $\pi$ transitions, respectively.

The two non-trivial (i.e. $\vert E \vert \ne 0$) solutions of eq.~(\ref{eq:wavematrix}) are found by setting the determinant of the matrix to zero, which results in a quadratic equation in $n^2$. Each of the two solutions $n_1$ and $n_2$ can then be substituted in to find a zero-value eigenvector $\vec{e}_{1,2}$ which together represent the (orthogonal) normal modes for propagation of light in the system, which are dependent on $\theta_B$.

To calculate the transmitted field, one must transform the $x,y,z$ basis into the normal mode basis, via the rotation matrix
\begin{align}
M = 
\left(
\begin{array}{ccc}
e_{11} & e_{12} & e_{13} \\
e_{21} & e_{22} & e_{23} \\
0 & 0 & 1
\end{array}
\right)^*,
\end{align}
where the $*$ denotes the complex conjugate, and after which propagation in the $z$-axis through a distance $L$ is computed by evolving each normal mode in space with its associated refractive index $n_1$ and $n_2$. This is done via the diagonal propagation matrix
\begin{align}
P = 
\left(
\begin{array}{ccc}
\exp({\rm i}(2\pi n_1 k L)) & 0 & 0 \\
0 & \exp({\rm i}(2\pi n_2 k L)) & 0 \\
0 & 0 & 1
\end{array}
\right).
\end{align}
Finally, one can use the inverse matrix $M^{-1}$ to transform back to the cartesian coordinates.

In terms of matrix operations, and including rotation around the $z$-axis, the full set of operations is
\begin{subequations}
\begin{align}
\vec{E}_{\rm out}(L) &= R_z(-\phi_B) \; M^{-1} \; P(L) \; M \; R_z(\phi_B) \vec{E}_{\rm in}, \\
 &= J_\chi \vec{E}_{\rm in},
\label{eq:JonesAtoms}
\end{align}
\end{subequations}
where for shorthand we combine all these processes into a single effective Jones matrix $J_\chi$~\cite{Jones1941}.

For arbitrary angle $\theta_B$, the solutions for $n_1$ and $n_2$ and their associated eigenvectors are not easy to write down analytically. In this case, we use the symbolic Python ({\tt sympy}) package to solve these equations. However, for the case where either $\theta_B=0,\pi$ (the Faraday geometry) or $\theta_B=\pi/2, 3\pi/2$, there are analytic solutions for $n_{1,2}$ and $\vec{e}_{1,2}$ which are much more computationally simple to implement. The Faraday geometry was the only case that the original version of ElecSus accounted for. In the next subsections we will describe these special cases in more detail.

\subsection{The Faraday geometry}

The Faraday geometry, where $\hat{k} \cdot \hat{B} = \pm 1$, is the geometry in which the Faraday effect [refs] is observed. In this geometry, we find the solutions for the refractive index are given by ( for $\theta_B = 0$)
\begin{align}
n_1 &= \sqrt{\epsilon_x - {\rm i}\epsilon_{xy}} = \sqrt{1 + \chi_+} \\
n_2 &= \sqrt{\epsilon_x + {\rm i}\epsilon_{xy}} = \sqrt{1 + \chi_-},
\end{align}
or alternately, that the two indices are associated with $\sigma^+$ and $\sigma^-$ transitions. 
The corresponding eigenvectors are
\begin{align}
\vec{e}_1 = \left(
\begin{array}{ccc}
{\rm i} \\ 1 \\ 0
\end{array} \right), \quad
\vec{e}_2 = \left(
\begin{array}{ccc}
-{\rm i} \\ 1 \\ 0
\end{array} \right).
\end{align}
It is clear in this case that applying the rotation matrix $M$ to the $x,y,z$ coordinate system simply transforms the coordinates into the circular basis, and we find as expected that $\sigma^{+}$ transitions couple directly to left circularly polarised light, and $\sigma^{-}$ transitions couple directly to right circularly polarised light. No component of the light couples to $\pi$ transitions in this geometry. Switching $\theta_B=0$ to $\theta_B=\pi$ simply inverts the coupling, i.e. $n_1$ and $n_2$ are swapped, resulting in the circular polarisations and $\sigma^\pm$ transitions coupling to each other in the opposite way.

In the general case, there is a different refractive index for each of the two circular polarisations (circular birefringence and dichroism), which on propagation leads to a rotation of the plane of polarisation (Faraday roatation).
Light that is initially linearly polarised can be decomposed into the cirular basis (in the $x-y$ plane), and on propagation if the medium is resonant with only one of the $\sigma^\pm$ transitions, one component of circular polarisation will be absorbed, leading to an effective circular polarisation filter - the output will be just the circular component that is not absorbed.

\subsection{The Voigt geometry}

The Voigt geometry, where $\hat{k} \cdot \hat{B} = 0$, is the geometry in which the magnetic field axis is transverse to the light propagation axis. The solutions for the refractive index in this geometry are~\cite{Rotondaro2015}
\begin{align}
n_1 &= \sqrt{\epsilon_x + \epsilon_{xy}^2/\epsilon_x} = \sqrt{\frac{2(1+\chi_+ + \chi_- + \chi_+ \chi_-)}{(2+\chi_+ + \chi_-)}} \\
n_2 &= \sqrt{\epsilon_z} = \sqrt{1 + \chi_0},
\end{align}
therefore $n_1$ is associated with both $\sigma^\pm$ transitions, while $n_2$ is associated with only $\pi$ transitions. 
The corresponding eigenvectors are (for $\theta_B = \pi/2$)
\begin{align}
\vec{e}_1 = \left(
\begin{array}{ccc}
0 \\ \epsilon_x / \epsilon_{xy} \\ 1
\end{array} \right), \quad
\vec{e}_2 = \left(
\begin{array}{ccc}
1 \\ 0 \\ 0
\end{array} \right).
\end{align}
For $\vec{e}_2$ it is clear that the field component that is parallel to the magnetic field (the $x$-axis in the case where $\phi_B=0$) drives $\pi$ transitions. The first eigenvector is harder to immediately visualise; we know that the eigenvectors must be perpendicular to both each other and $k$, and hence $\vec{e}_1$ must point along the cartesian axis perpendicular to both $k$ and $B$ (i.e. it points along $y$ in the case where $\phi_B=0$). The normal mode is elliptically polarised in the plane perpendicular to $\vec{B}$, i.e. in the $y-z$ plane. Since the atomic quantisation axis lies along $\vec{B}$, a linearly polarised beam along $y$ can be decomposed, in the atomic frame, into equal circular components in the $y-z$ plane, and thus it drives (equally) both $\sigma^+$ and $\sigma^-$ transitions. 

In contrast to the Faraday case where the medium exhibits {\it circular} birefringence and dichroism, in the Voigt geometry the system is {\it linearly} dichroic and birefringent, which leads to a different form of magneto-optic rotation known as the Voigt effect~\cite{Voigt1898,Schuller1991,Budker2002}. Note that, as pointed out by Pershan~\cite{Pershan1967}, the Voigt effect is subtly different to the Cotton-Mouton effect~\cite{Cotton1905,Beams1932}, where birefringence emerges as a result of alignment of diamagnetic molecules in a transverse magnetic field.

\section{Jones Matrices for propagating fields}

In the previous section, the use of matrices is a simple and computationally convenient method for calculating the propagation of the electric field in the medium. These matrix methods are generally referred to as Jones calculus~\cite{Jones1941}, and similar matrices can be constructed for a variety of common optical elements, including waveplates and polarisers (see table~\ref{tab:jones} in the appendix). 
Using these matrices, one can calculate the output electric field (and intensity) from any arbitrary combination of optics.

\subsection{Stokes parameters using Jones Matrices}

As in the previous version, the four Stokes parameters $S_{0,1,2,3}$ are an easily measurable way of characterising the polarisation state of light. The Jones calculus approach offers an intuitive way of calculating these parameters, and only require calculation of the output field. 
The four Stokes parameters are
\begin{subequations}
\begin{align}
S_0 &\equiv (I_{\rm LCP}+I_{\rm RCP})/I_0 = (I_x+I_y)/I_0 = (I_\nearrow+I_\searrow)/I_0, \label{eq:S0def} \\
S_1 &\equiv (I_x-I_y)/I_0, \label{eq:S1def} \\
S_2 &\equiv (I_\nearrow - I_\searrow)/I_0, \\
S_3 &\equiv (I_{\rm RCP} - I_{\rm LCP}) / I_0.
\label{eq:Stokes}
\end{align}
\end{subequations}
In the above equations, $S_0$ is the normalised output intensity and requires no extra matrices - all that is required is to sum the mod-squared field components ($I_j = \varepsilon_0 c \vert E_j \vert^2$/2, with $j x,y,{\rm LCP}, {\rm RCP}$) in whichever orthogonal basis is most appropriate.
Each of the components $E_x, E_y, E_{\rm LCP}, E_{\rm RCP}, E_\nearrow$ and $E_\searrow$ can be found by applying the respective Jones matrix to the output field - for example,
\begin{align}
E_x = \hat{\mathcal{J}}_x E_{\rm out}, \quad E_y = \hat{\mathcal{J}}_y E_{\rm out},
\end{align}
which are required to compute the $S_1$ parameter, the difference between the linear basis components in the $x$ and $y$ plane.
$S_2$ is the difference between orthogonal linear polarisations at 45 degrees to the $x$ and $y$ axes, and $S_3$ is the difference between the two circular polarisation components (note the change in notation from the previous publication - in a non-Faraday geometry, $I_+ \ne I_{\rm L(R)CP}$).

\section{New applications of ElecSus}
\label{sec:Further}

In this section we illustrate the new features of ElecSus through a series of example data sets. Note that the examples here are all for Rb, but ElecSus will work with the alkali-metal atoms Rb, Cs, K and Na.

\begin{figure}[t]
	\begin{center}
	\includegraphics[width=0.98\columnwidth]{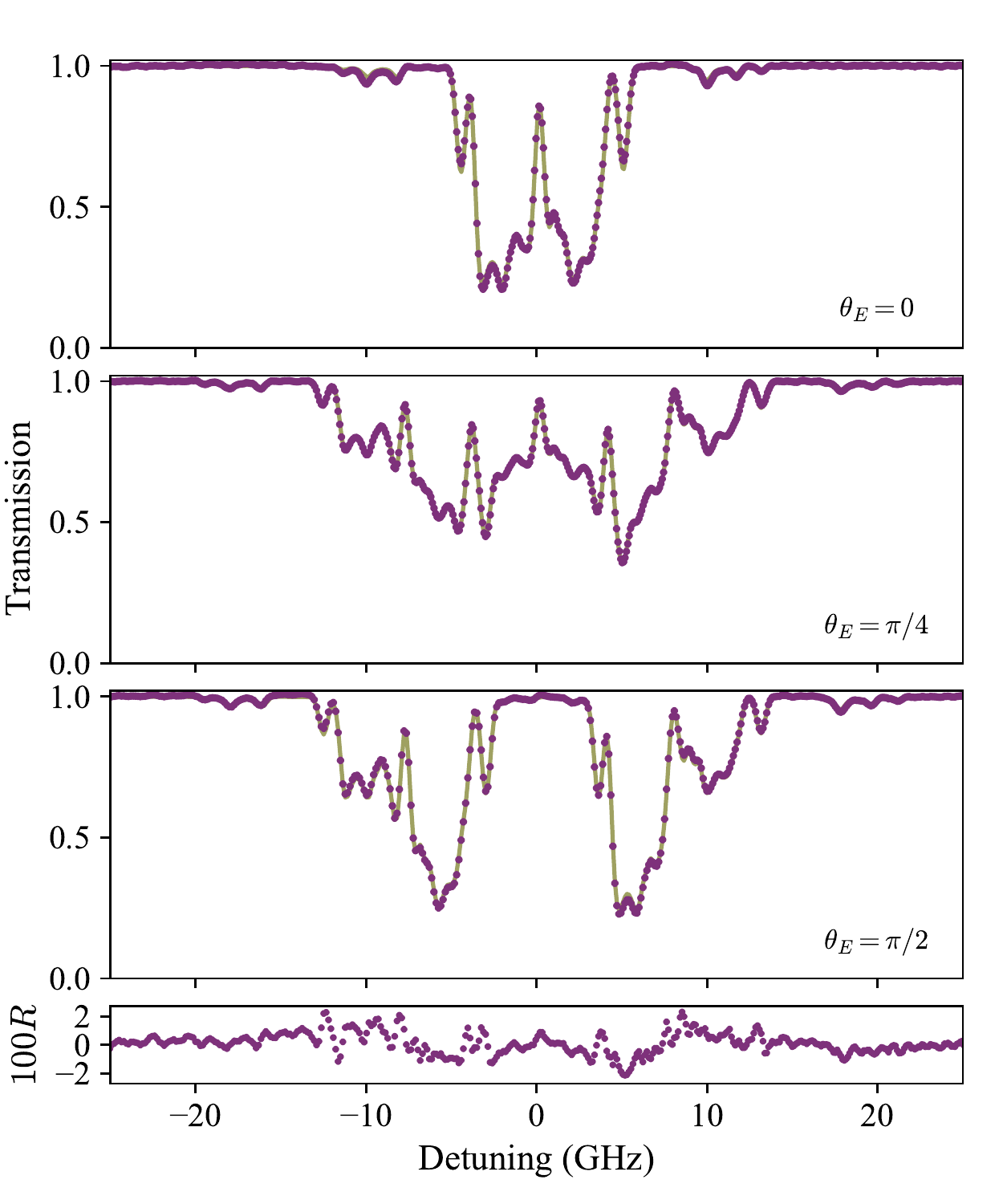}
	\end{center}
	\caption{Transmission spectra through a 1~mm naturally abundant Rb vapour cell in the Voigt geometry for 3 incident linear polarisation angles. Data are shown as purple points, the ElecSus fits are shown as olive lines. For each data set, we fit the applied magnetic field, cell temperature and incident polarisation angle (we assume linear polarisation). The top panel shows the case for $\theta_E=0$, where only $\pi$ transitions are driven. The second panel shows $\theta_E = \pi/4$, where $\pi$ and $\sigma^\pm$ transitions are driven equally. The third panel shows $\theta_E = \pi/2$, where only $\sigma^\pm$ transitions are driven. Finally, the bottom panel shows the residuals between experiment and theory (amplified by a factor of 100) for the third panel, clearly indicating an excellent fit. Fit parameters are temperature, magnetic field and incident polarisation angle.}
	\label{fig:VoigtS0}
\end{figure}

\subsection{Transmission spectroscopy in the Voigt geometry}

Transmission spectroscopy in the Voigt geometry is the simplest method to demonstrate the addition of $\pi$ transitions to ElecSus.
%
%
We perform weak-probe~\cite{Sherlock2009} transmission spectroscopy (intensity approximately 0.03 mW/cm$^2$)
through a 1 mm naturally abundant Rb vapour cell. The cell temperature was approximately 95$^\circ$C, and the applied magnetic field strength was approximately 0.42~T, along the $x$-axis. 
The laser source is a distributed feedback (DFB) laser (quoted linewidth approx. 2~MHz), which is frequency tuned by temperature tuning of the diode. This allows a mode-hop-free laser scan over an extremely large detuning range (up to $\sim 1$~THz). The scan is linearised with a Fabry-Perot etalon, and a room temperature 75~mm reference cell is used as an absolute frequency reference, following the procedure outlined in reference~\cite{Keaveney2013}. Tuning the diode temperature slightly changes the laser output power, so we stabilise the optical power that is incident on the atomic vapour using a feedback loop linked to the RF power of an AOM, as described in ref.~\cite{Truong2015a}.

Figure~\ref{fig:VoigtS0} shows the resulting transmisison spectra for 3 different angles of incident linear polarisation. For a linearly polarised input electric field, we define $\theta_E$ as the angle the electric field makes with the $x$-axis for simplicity. Data are shown as purple points. For light polarised with the electric field along $x$ ($\theta_E = 0$, top panel), $E \parallel B$ and therefore the only allowed transitions are $\pi$ transitions. When the electric field oscillates in the $y$-axis ($\theta_E = \pi/2$, third panel), $E \perp B$ and hence $\sigma^\pm$ transitions are driven. Finally, when the electric field is at 45 degrees to the $x,y$ axes (second panel), all three types of transition are driven. At this magnetic field strength, for naturally abundant Rb, the spectra in all three cases are very complex. However, we see in all three panels the fit to the data using ElecSus, which are in excellent agreement in all 3 cases (the RMS errors are 0.64\%, 0.36\% and 0.59\% for the $\theta_E = 0, \pi/4, \pi/2$ data, respectively). We also plot the residuals (difference between theory and experiment) for the $\theta_E = \pi/2$ data, multiplied by a factor of 100, on the bottom panel; the lack of any clear structure in the residuals indicates that the theoretical model incorporates all of the underlying physics.

\subsection{Stokes polarimetry in the Voigt geometry}

%
\begin{figure*}[t]
	\begin{center}
	\includegraphics[width=1.5\columnwidth]{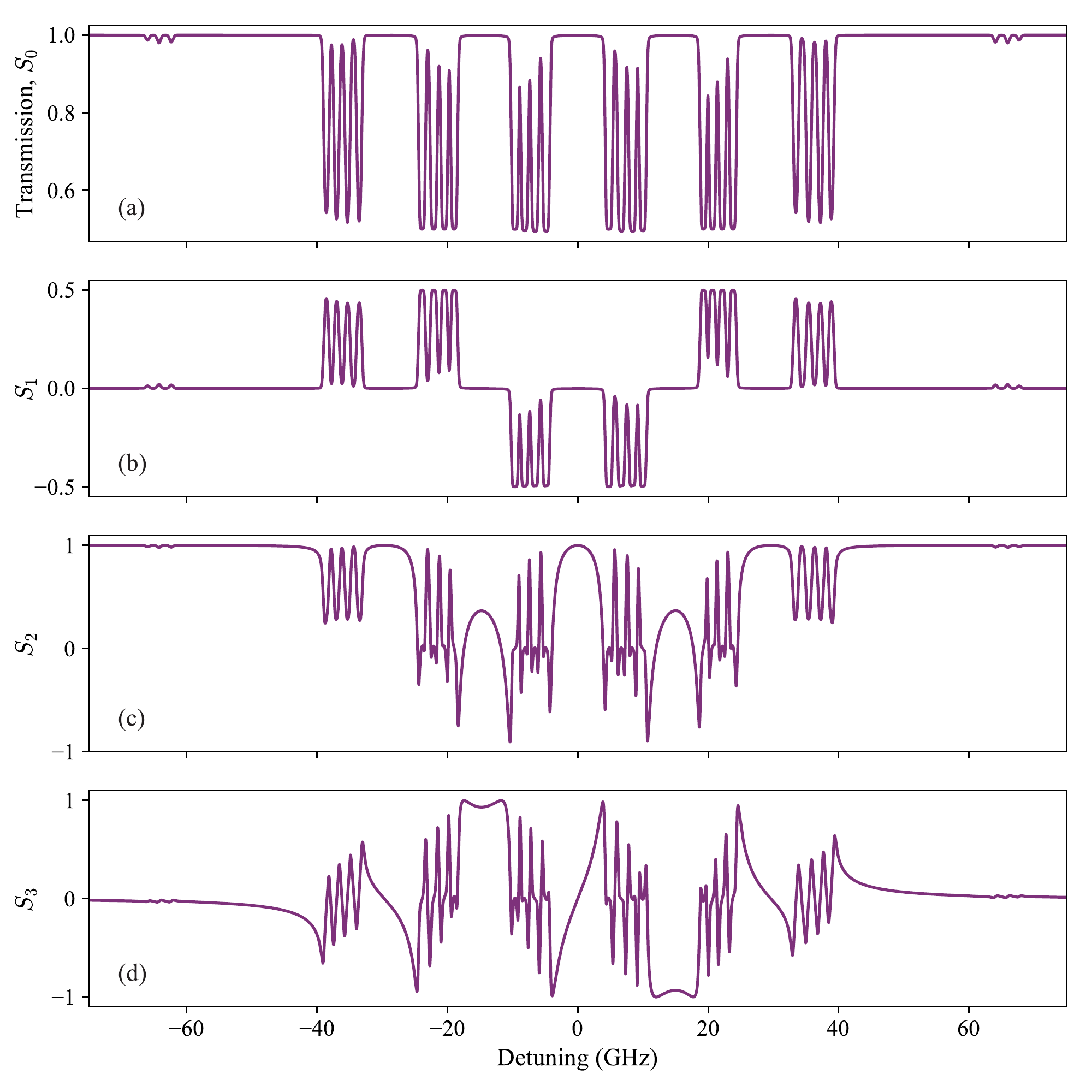}
	\end{center}
	\caption{Theoretical prediction of the Stokes parameters in the Voigt geometry at $B=1.54$~T, $T=125^\circ$C, $L=1$~mm, in a isotopically enriched 99\% $^{87}$Rb vapour on the D2 line. The input electric field is linearly polarised, with $\theta_E = \pi/4$.}
	\label{fig:VoigtS0123}
\end{figure*}

In the absence of an applied magnetic field, or in a small applied field, Doppler broadening masks the complex atomic structure in a thermal vapour. Is is therefore more instructive to consider the case of a large applied field. This regime, known as the hyperfine Paschen Back (HPB) regime, has generated much recent interest~\cite{Olsen2011,Weller2012,Weller2012a,Weller2012b,Sargsyan2014,Zentile2014a,Sargsyan2015b,Sargsyan2017a,Ma2017a}. 
For $^{87}$Rb in the HPB regime, due to the large Zeeman splitting, the atomic transitions are all separated by more than the width of the Doppler broadened lines. 
For some applications with thermal vapours, this greatly simplifies the physics, since true 2-, 3-, and 4-level systems can be easily isolated, allowing for archetypal demonstrations of selective reflection~\cite{Sargsyan2017}, electromagnetically induced transparency~\cite{Whiting2016a}, electromagnetically induced absorption~\cite{Whiting2015}, and four-wave-mixing processes~\cite{Whiting2017,Whiting2017a}.

In figure~\ref{fig:VoigtS0123} we plot theoretical predictions of the Stokes parameters for a Rb D2 line spectrum (isotopically enriched, 99\% $^{87}$Rb) with an applied magnetic field of 1.54~T, in the voigt geometry ($\theta_B = \pi/2, \phi_B=0$). The vapour cell length is 1~mm, cell temperature is 120~$^\circ$C and the input polarisation is linear, with $\theta_E = \pi/4$ ($\vec{E}_{\rm in} = \vec{E}_\nearrow = 1/\sqrt{2} \; [1,1,0]^{\rm T}\vert E_{\rm in} \vert$, where the T denotes the matrix transpose). 
The choice of magnetic field, vapour cell length and isotopic composition in figure~\ref{fig:VoigtS0123} was chosen to match an experimental setup available to us, which will be described later.

From the $S_0$ spectrum (fig.~\ref{fig:VoigtS0123}(a)), we can observe 6 main (strong) sets of features in groups of 4, and two sets of visible smaller (weak) features on the far edges of the spectrum. The group of 4 comes from the projection of the nuclear spin quantum number $m_I$, which for $^{87}$Rb (nuclear spin $I=3/2$) can take four values: -3/2, -1/2, 1/2 and 3/2. The two inner-most groups of four originate from $\pi$ transitions ($m_J = \pm 1/2 \rightarrow m_J' = \pm 1/2$), whilst the outer groups are from $\sigma^-$ transitions on the side of negative detuning ($m_J = 1/2 \rightarrow m_J' = -1/2$, and $m_J = -1/2 \rightarrow m_J'=-3/2$) and $\sigma^+$ transitions on the side of positive detuning ($m_J = 1/2 \rightarrow m_J' = 3/2$, and $m_J = -1/2 \rightarrow m_J'=1/2$).
The weak transitions at $\sim\pm60$~GHz stem from the incomplete decoupling of the ground state $5S_{1/2}$ into the $\vert m_S, m_I \rangle$ basis ($m_L = 0$ for the all terms in the ground state manifold) - the ground states are not pure eigenstates in this basis and there is therefore a small admixture of other states which results in the weak transitions, as described in ref.~\cite{Zentile2014a} for the Faraday geometry.

The $S_1$ spectrum (fig.~\ref{fig:VoigtS0123}(b)) shows the consequence of the linear dichroism of the medium. Off resonance, there is no interaction and hence $I_x = I_y = I_0 / 2$ and $S_1 = 0$. At the atomic resonance frequencies, one component of light in the linear basis is completely absorbed, leading to either $I_{x,y} = 0$ while the other component far off-resonance and therefore unaffected. The $S_1$ spectrum therefore swings between values of $\pm 0.5$ depending on which component is absorbed.

The $S_2$ spectrum (fig.~\ref{fig:VoigtS0123}(c)) is the difference between the linear polarisation components but in the diagonal ($I_\nearrow - I_\searrow$) basis. This spectrum therefore has an off-resonance value of $1$ since we input the $\nearrow$ polarisation. On resonance the absorption dominates, which reduces both $I_\nearrow$ and $I_\searrow$ and the $S_2$ value tends towards zero. Between the resonances, however, there is still optical rotation (since neither $\nearrow$ nor $\searrow$ are eigenmodes of propagation); this is most pronounced between the $\pi$ and $\sigma^\pm$ groups at $\pm15$~GHz detuning since the optical rotation adds.

Finally, the $S_3$ spectrum (fig.~\ref{fig:VoigtS0123}(d)) is the Voigt-geometry equivalent of a Faraday rotation spectrum (which are observed in the $S_{1,2}$ Stokes parameters), which is intuitively what one might expect since in the Faraday case the medium is circularly birefringent and the polarisation rotation is observed in the linear basis, whilst the Voigt case is the opposite way around - the medium is linearly birefringent and the rotation can therefore be observed in the circular basis.

\subsection{Modelling cell window birefringence}

%
\begin{figure}[t]
	\begin{center}
	\includegraphics[width=0.8\columnwidth]{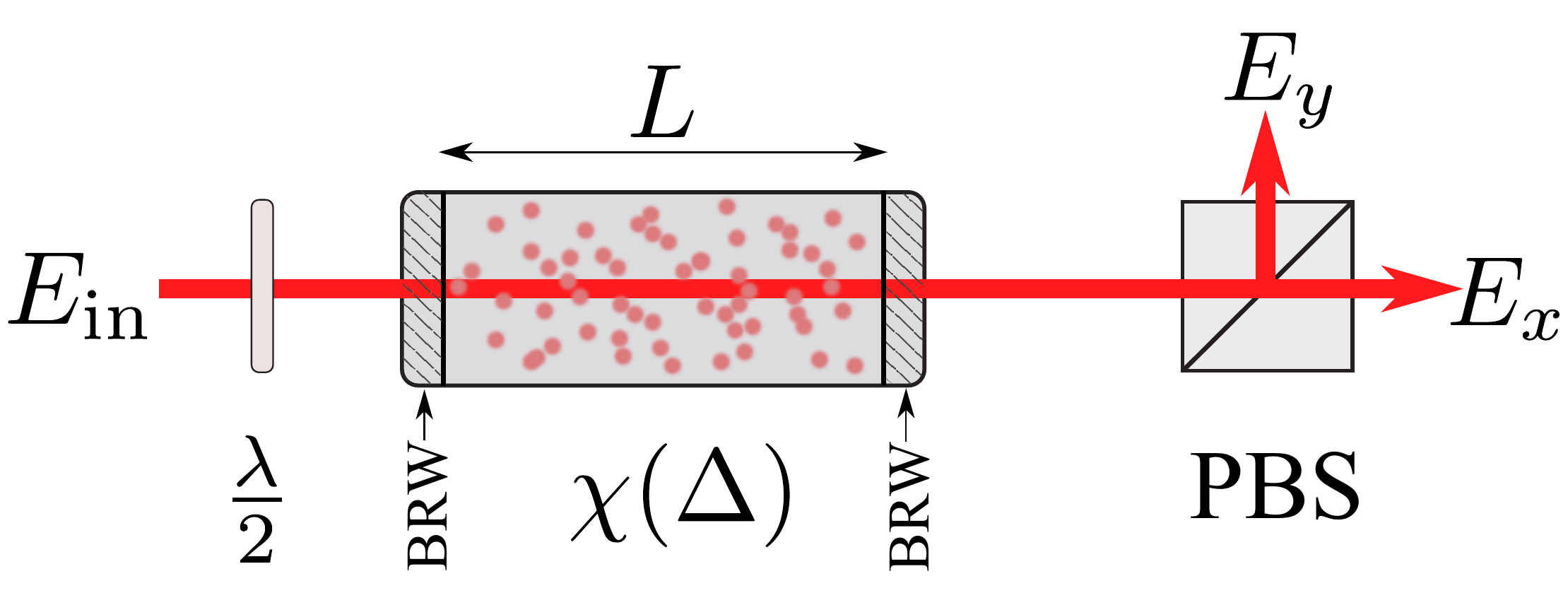}
	\end{center}
	\caption{Electric field propagation example. Each optical element, including the vapour cell, can be described by a Jones matrix which can be cascaded to calculate the output of any series of optical elements. This example shows a typical experimental setup to measure the $S_1$ Stokes parameter: the initial input polarisation $E_{\rm in}$ is rotated by the half-waveplate ($\lambda/2$), passes through the cell including the birefringent windows (BRW), after which the field components $E_x$ and $E_y$ are analysed by the transmission and reflection (respectively) through a polarising beam splitter cube (PBS).}
	\label{fig:Efield}
\end{figure}

An issue in experimental polarimetry measurements comes from birefringence in optical elements, which causes unwanted additional optical rotation. 
In thermal vapour experiments, a common source of this unwanted birefringence is the vapour cell windows, but the amount of birefringence is not usually known a priori. We can model the effect of an unknown birefringent material through a Jones matrix, $J_{\rm BRW}(\phi_{\rm BR}, \theta_{\rm BR})$ (see Appendix for details)
which is characterised by 2 parameters, the phase shift $\phi_{\rm BR}$ and the orientation with respect to the optical axis $\theta_{\rm BR}$ (the subscript BR is used to differentiate these two parameters from the magnetic field angles used earlier).

Figure~\ref{fig:Efield} illustrates an example situation. An input electric field $E_{\rm in}$ travels through a half waveplate ($\lambda/2$), then through a vapour cell containing an atomic medium of length $L$. The vapour cell has birefringent windows (BRW) on both ends. After the vapour cell, the polarisation state is analysed by placing a 
polarising beamsplitter cube (PBS) which analyses the $x$ and $y$ components of the electric field. 
The Jones calculus approach is to cascade the matrices for all these optical elements. 
Combining all elements, the outputs $\vec{E}_{x,y}$ after the PBS cube would be
\begin{align}
\vec{E}_{x,y} = J_{x,y} J_{\rm BRW}(\theta_{\rm BR}, \phi_{\rm BR}) J_{\rm atoms} J_{\rm BRW}(\theta_{\rm BR}, \phi_{\rm BR}) J_{\lambda/2}(\theta_{\rm H}) \vec{E}_{\rm in}.
\end{align} 
Taking the difference in intensity between the two output ports of the beam splitter, we measure the $S_1$ Stokes parameter.

As a demonstration of the effects of window birefringence, we now show some experimental polarimetry data. Our optical setup is the same as shown in figure~\ref{fig:Efield}. The applied magnetic field is produced by a `magic sphere' configuration of NdFeB permanent magnets~\cite{Trenec2011} which yields a peak field of 1.54~T at the centre of the hollow cylinder, where we place a microfabricated vapour cell~\cite{Liew2004} inside a small copper heating block. Right-angled prisms allow the light to propagate through the cell at normal incidence to the field, realising the Voigt geometry.
\begin{figure}[t]
	\begin{center}
	\includegraphics[width=0.98\columnwidth]{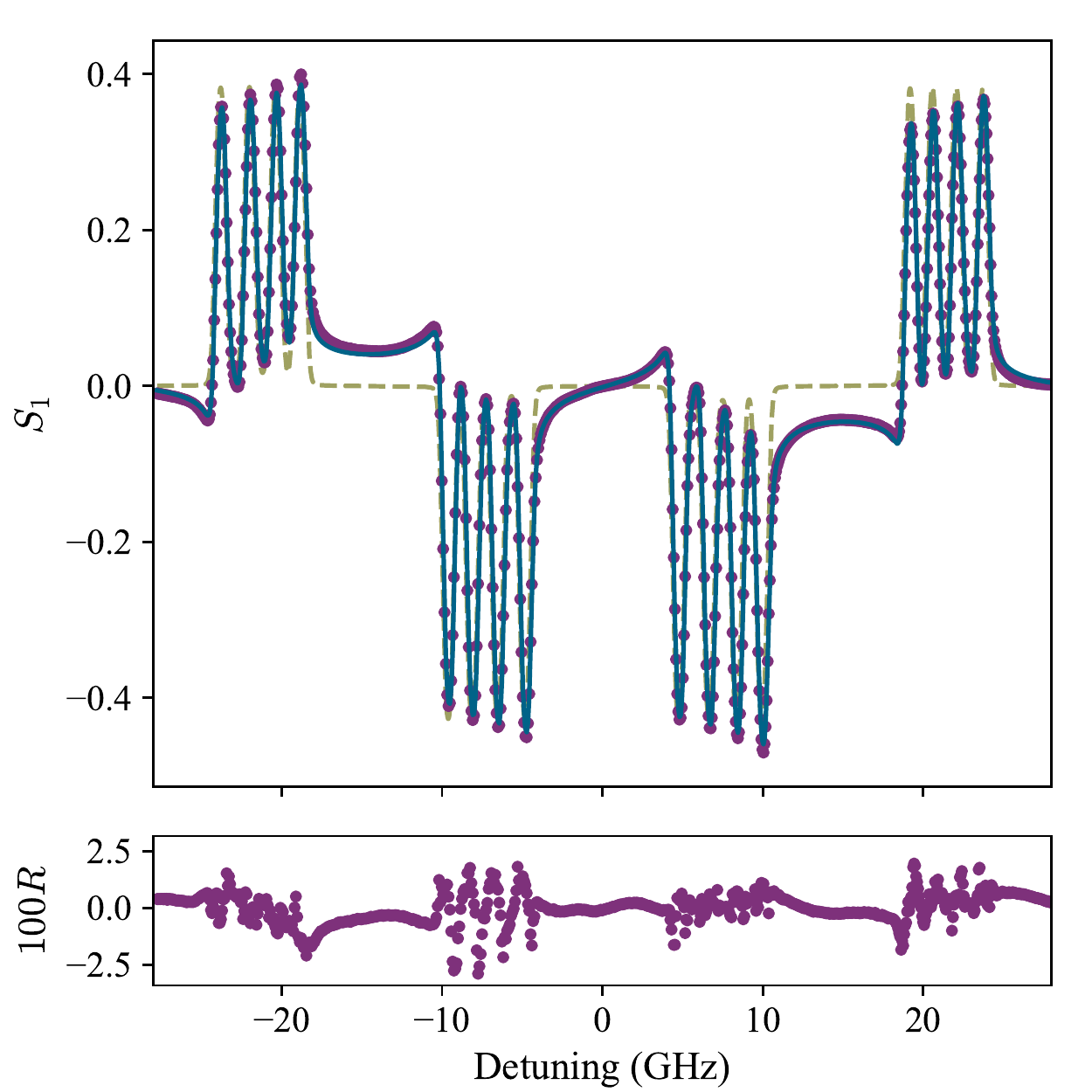}
	\end{center}
	\caption{Example of the effect of cell window birefringence on optical rotation spectra. Purple points show experimental data from a 1~mm isotopically enriched (99\% $^{87}$Rb) vapour cell in an applied magnetic field of 1.54~T and cell temperature $98^\circ$C. The olive dashed line shows the theoretical prediction without cell window birefringence. The blue line is the result of a fit, which allows the input linear polarisation angle and cell window birefringence parameters (phase shift and alignment of the optical axis) to vary. We assume that both cell windows have the same birefringent properties. Including birefringence clearly improves the agreement between experiment and theory, as shown by the small residuals on the bottom panel.}
	\label{fig:VoigtS1}
\end{figure}

The $S_1$ spectroscopic data are shown in figure~\ref{fig:VoigtS1}. The experimental conditions are similar to those of figure~\ref{fig:VoigtS0123}, except the cell temperature which is 98$^\circ$C.
The input polarisation is set so that far off-resonance, the difference signal is zero. In the absence of window birefringence, this would mean an input polarisation angle $\theta_E = \pm \pi/4$. However, when including window birefringence, the input polarisation needs to be rotated by a small amount to satisfy this condition.
\begin{figure*}[t]
	\begin{center}
	\includegraphics[width=2\columnwidth]{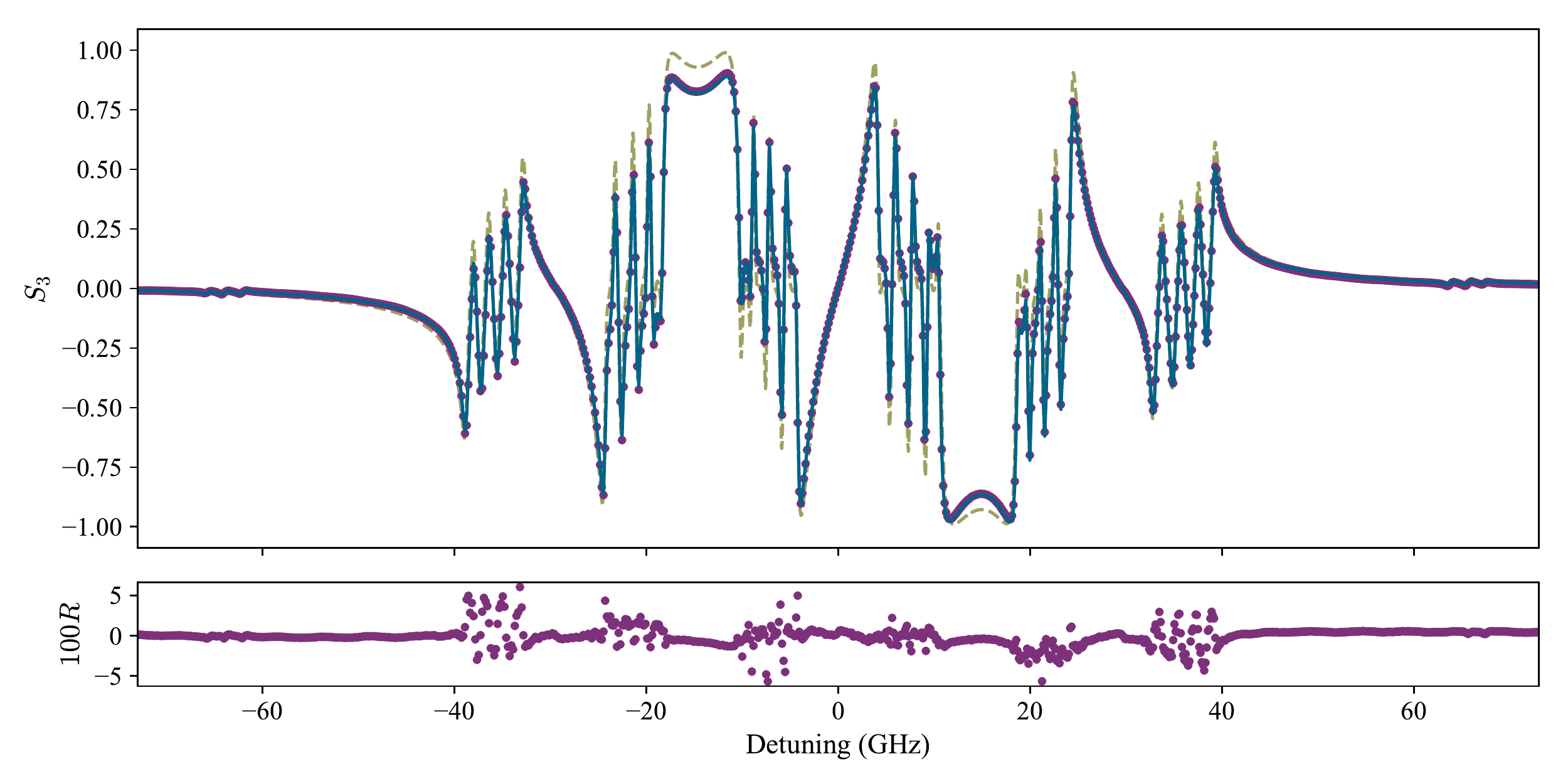}
	\end{center}
	\caption{Experimental S3 data with fit to theory, including an extra half-waveplate to compensate for cell window birefringence. Experimental conditions are the same as for figure~\ref{fig:VoigtS0123}. Purple points are experimental data, the dashed olive line is the expected signal in the absence of birefringence, and the solid blue line is a fit to the data taking into account window birefringence and a compensating half-waveplate.} 
	\label{fig:VoigtS3}
\end{figure*}
The data (purple points) show the main optical rotation features displayed in the $S_1$ spectrum of figure~\ref{fig:VoigtS0123}, except in the wings of the resonance lines. This is most prominent at around $\pm15$~GHz detuning, where the birefringence of the cell windows causes an additional rotation.
On the figure we plot two theoretical curves. The dashed olive line is the theory without birefringence, whilst the blue solid line is the result of a fit to the data, assuming both windows have equal birefringent properties, and allowing the birefringence parameters $\phi_{\rm BR}$, $\theta_{\rm BR}$ and the input linear polarisation angle $\theta_E$ to vary, along with the cell temperature and magnetic field strength. We find excellent agreement with the experimental data, as demonstrated by the small residuals on the bottom panel of figure~\ref{fig:VoigtS1} - the RMS error between theory and experiment is 0.8\%. This may therefore be a useful technique to practically determine the birefringent properties of such windows.

In some cases the birefringence can be compensated for with the addition of waveplates to the optical setup. 
In figure~\ref{fig:VoigtS3} we show an experimental $S_3$ spectrum, with the same conditions as figure~\ref{fig:VoigtS1} apart from the cell temperature, which is 125$^\circ$C. The optical setup is similar to figure~\ref{fig:Efield}, with the addition of a half-waveplate and a quarter-waveplate between the vapour cell and the PBS cube. The quarter waveplate and the PBS constitute a circular polarisation analyser, which is the usual experimental technique for measuring $S_3$. The extra half-waveplate is used to compensate for the cell birefringence which would otherwise offset the rotation signal, resulting in a spectrum that is qualitatively very similar to that of figure~\ref{fig:VoigtS0123} (the expected spectrum without birefringence is the olive dashed line in figure~\ref{fig:VoigtS3} for direct comparison). 

For the fit to this dataset, we constrain the cell birefringence properties to be those from the fit in figure~\ref{fig:VoigtS1}, and instead allow the angle of the half- and quarter-waveplates after the cell to vary.
In this case, we again find excellent agreement between the data and the fit, with RMS residuals of 1.1\%.

\subsection{Arbitrary angle geometry spectroscopy}

For the most general case where $\theta_B \ne 0, \pi/2$, the spectral features are very rich. In a study of magneto-optic filtering, Rotondaro, Zhdanov and Knize~\cite{Rotondaro2015} showed that the bandwidth of atomic filters can be reduced by using a non-standard magnetic field geometry. Again, here we present a comparison of ElecSus to an experimental data set to illustrate its use in these situations.
The experimental setup utilises a 1~mm natural abundance Rb cell, with an applied magnetic field provided by two permanent top-hat shaped magnets set up on a rotation platform, such that a range of angles can be formed between the light propagation axis and the magnetic field axis, limited only by the radial extent of the magnets and their mechanical mounts. The maximum field strength is limited to around 0.4~T with this setup. 

\begin{figure}[t]
	\begin{center}
	\includegraphics[width=0.98\columnwidth]{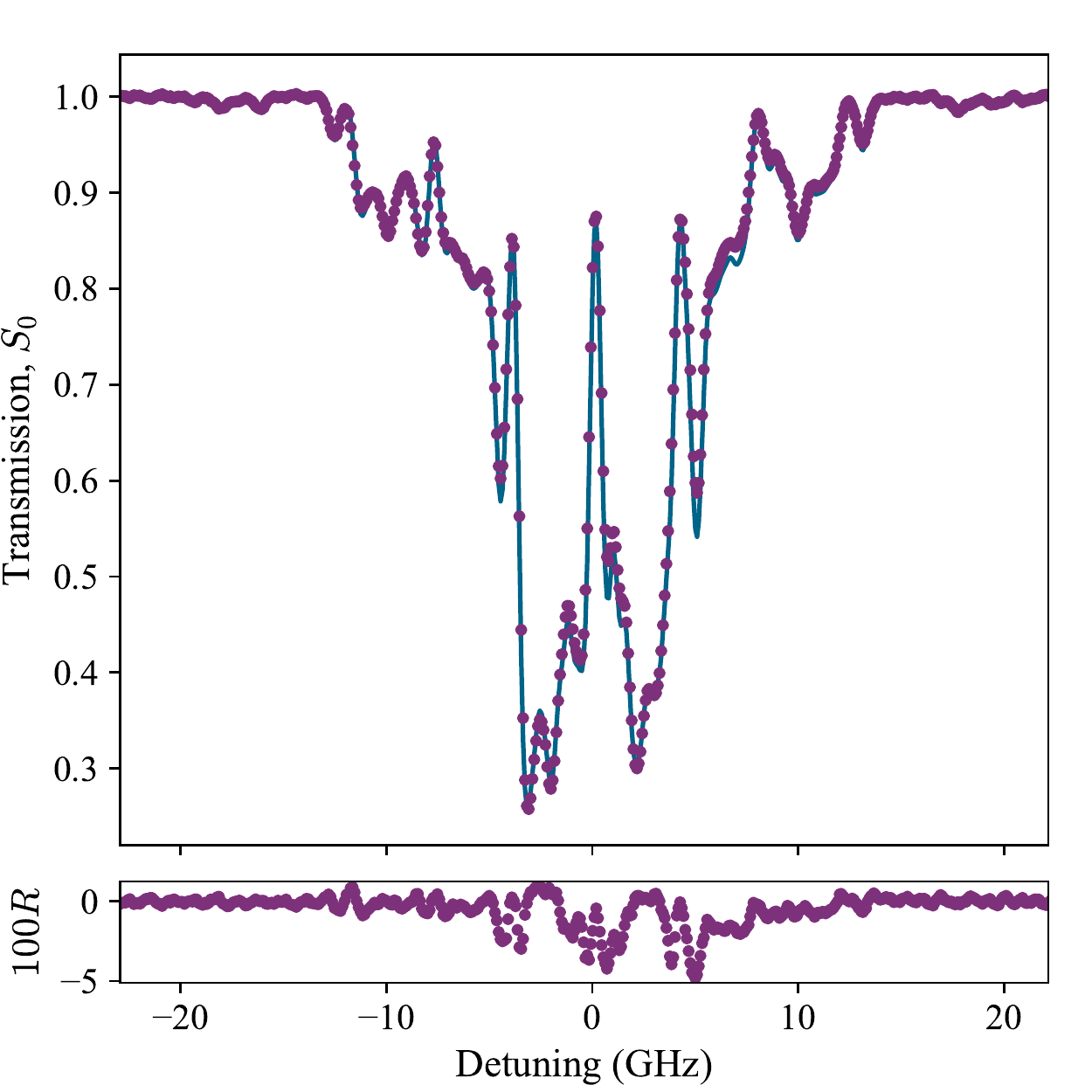}
	\end{center}
	\caption{Example of arbitrary angle geometry spectroscopy. The magnetic field (strength $B = 0.42$~T) is oriented at an angle $\theta_B \approx \pi/3$ from the $z$-axis, and we take an $S_0$ spectrum through a 1~mm naturally abundant Rb cell at a temperature $T=90^\circ$C, with linearly polarised input light ($\theta_E \approx 0$). Some similarity can be noted between this data and that of figure~\ref{fig:VoigtS0} (the magnetic field strengths are approximately equal), but there are clear differences, noably around $\pm7$~GHz, from the Voigt geometry with the same input polarisation. The purple points are experimental data, and the blue line is a fit to the data using ElecSus, allowing $\theta_E, \theta_B, B$ and $T$ to vary. } 
	\label{fig:arbangle}
\end{figure}

In figure~\ref{fig:arbangle} we show an $S_0$ spectrum with a magnetic field strength of $0.42$~T at an angle $\theta_B \approx \pi/3$. The experimental data are shown as purple points, and the fit to the data using ElecSus is shown as a blue line. Clearly, experiment and theory match very well, as shown by the residuals. The RMS error between theory and experiment is 0.8\%. Since the field strength is similar to the data in figure~\ref{fig:VoigtS0}, the atomic resonance positions are in nearly the same place. We can then compare for similar conditions (i.e. incident polarisation angle) which transitions are driven. The central features within approximately $\pm 7$~GHz originate from $\pi$ transitions, so there is similarity to the top panel of fig.~\ref{fig:VoigtS0}. However, the features at larger detuning in fig.~\ref{fig:arbangle} come from $\sigma^\pm$ transitions and are therefore not present when interrogating the medium with $\theta_E=0$ polarised light in the Voigt geometry. The relative strength of the atom-light coupling is also completely different to any of the Voigt-geometry cases; for the data in figure~\ref{fig:arbangle} the $\pi$ transitions are clearly more strongly driven than the $\sigma^\pm$ transitions.

This data set demonstrates that ElecSus is now suitable for use with arbitrary angle magnetic fields. We also note that we can succesfully reproduce the plots from reference~\cite{Rotondaro2015}; we provide a Python script to reproduce these figures in the \texttt{tests/} subdirectory of the GitHub repository.

\subsection{Magnetic field gradients}

\begin{figure}[t]
	\begin{center}
	\includegraphics[width=0.98\columnwidth]{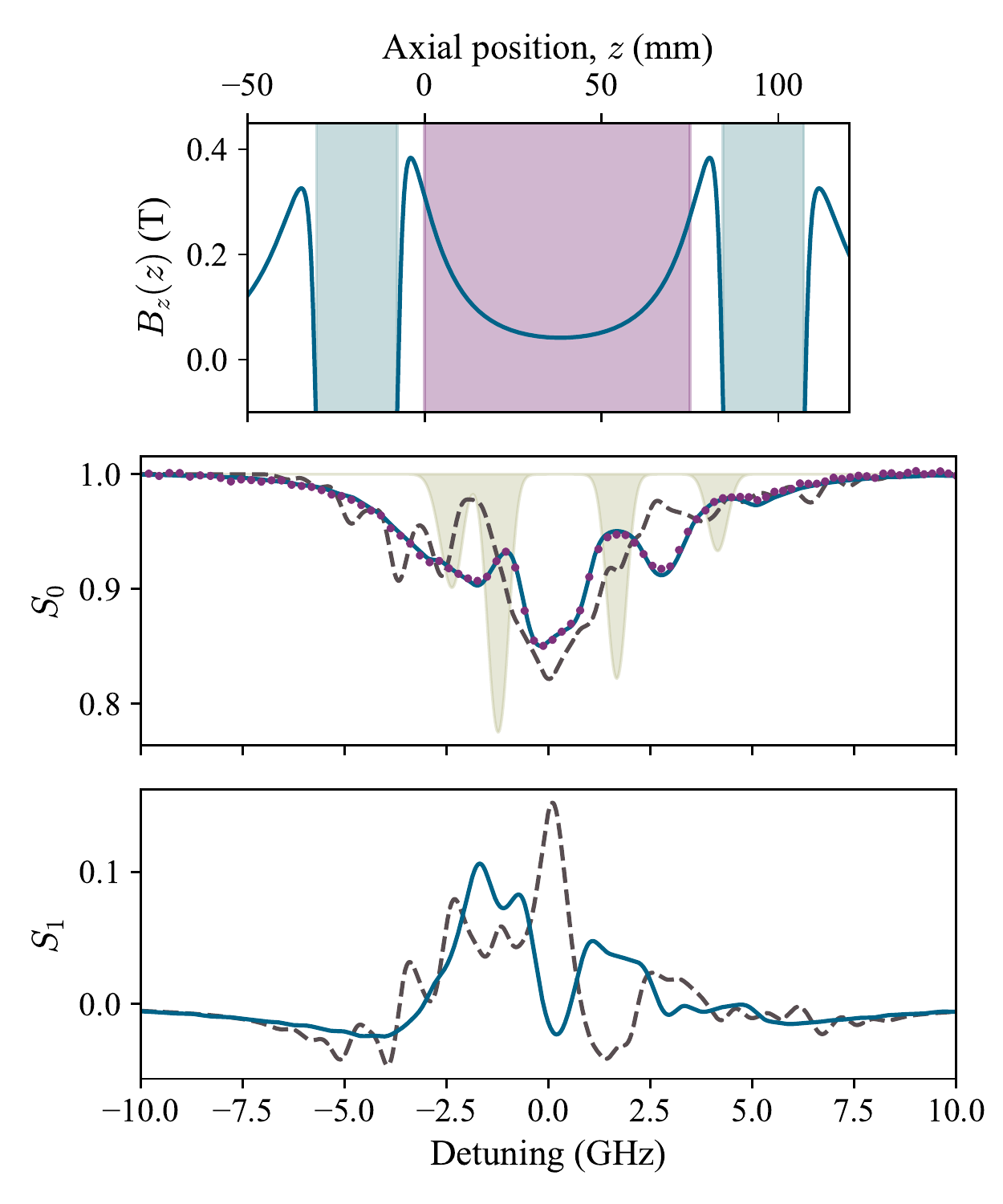}
	\end{center}
	\caption{Example of spectroscopy with a magnetic field gradient. Top panel shows magnetic field profile between two top-hat magnets (axial extent marked by blue shading), separated by 92 mm with a 75~mm vapour cell placed between them (purple shading). Middle and bottom panels show $S_0$ and $S_1$ signals from a room temperature, naturally abundant 75~mm long Rb vapour as the probe laser is scanned across the D2 line. The olive shading shows the expected resonance positions in the absence of an applied magnetic field (but with scaling altered for clarity). The purple points are experimental data. The grey dashed lines are calculations of $S_{0,1}$ which assume the mean value of magnetic field ($B_{\rm avg} \approx 1$~kG) across the cell, while the solid blue line is a fit to the data using the full field profile $B_z(z)$. The only fit parameters are the position of the two magnets (their spacing and offset relative to the cell position) and the temperature of the vapour.}
	\label{fig:gradient}
\end{figure}

Since the electric field is now calculated explicitly, and can be returned directly by the program, it is possible to use ElecSus to now predict spectra from non-uniform systems - this could be, for example, 
the magnetic field gradient across a thermal vapour cell.
We place a room temperature, 75~mm naturally abundant Rb vapour cell between two top-hat magnets, which are separated by a little more than the vapour cell length. This creates an axial magnetic field profile (the Fardaay geometry) shown in the top panel of figure~\ref{fig:gradient}, which has calculated minimum/maximum/mean values inside the vapour cell of 41~mT,  311~mT and 100~mT, respectively. The middle and bottom panels show transmission spectroscopy and the Faraday rotation signal $S_1$ in this experimental configuration. Purple points are experimental data. The dashed grey lines show a calculation which assumes the mean value of magnetic field, whilst the blue lines show a calculation which splits the cell into, in this case, 25 segments, and propagates the electric field through each segment sequentially. The two models are very clearly different in both $S_0$ and $S_1$ spectra. We fit using this calculation; the fit parameters are the vapour cell temperature, and the position of the two top-hat magnets (relative separation and position offset with respect to the cell; the magnets' remnant field strength is fixed). There is excellent agreement between this model and experimental data (the RMS error is 0.2\%).

\subsubsection{Faraday filter with field gradient}

Though the previous example showed the case of an extreme field gradient, it is not likely to be a practically relevant case.
In this subsection we consider the application of field gradients to Faraday filtering.
\begin{figure}[t]
	\begin{center}
	\includegraphics[width=0.9\columnwidth]{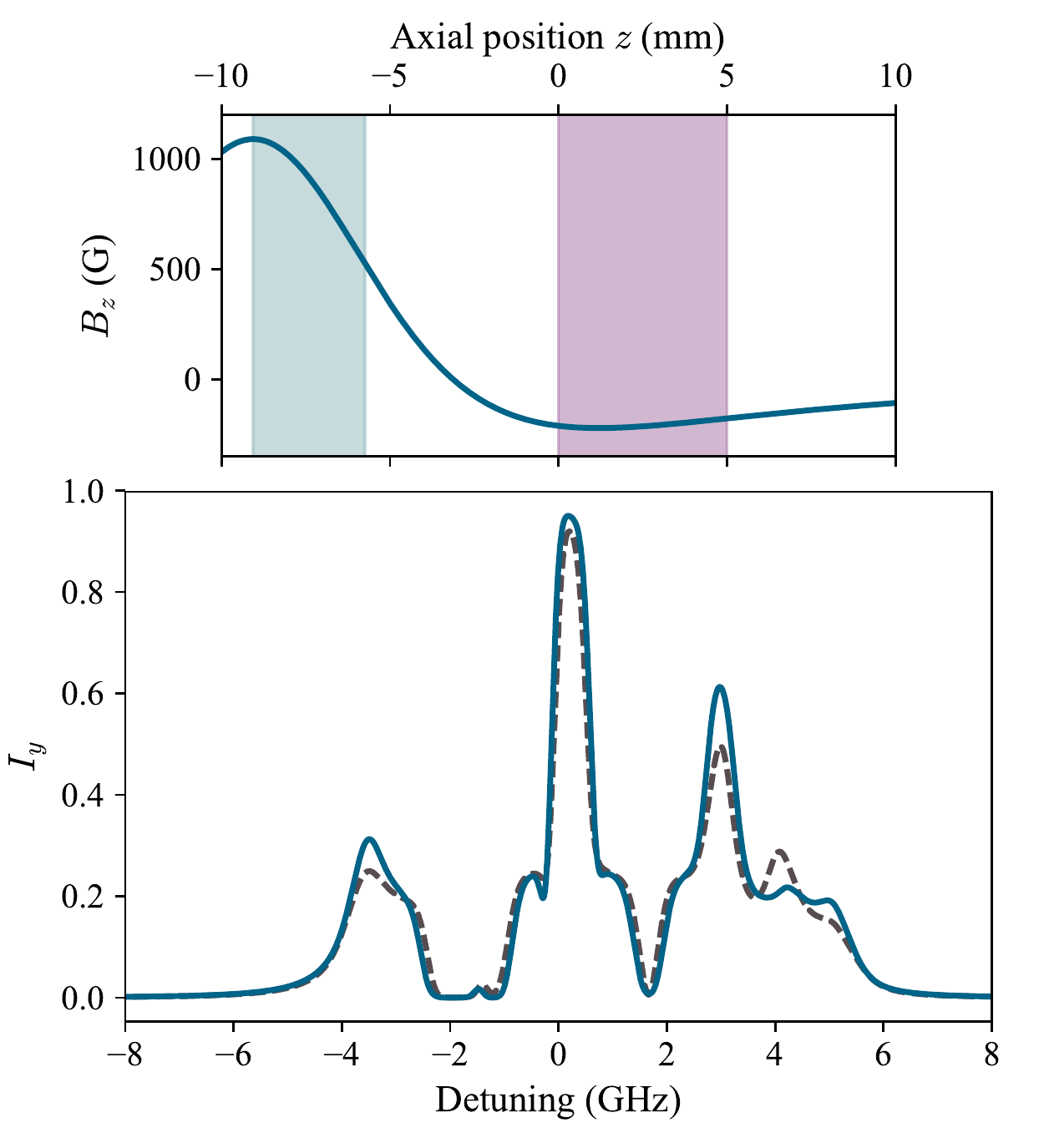}
	\end{center}
	\caption{Comparison of a Faraday filter with non-uniform and uniform fields. The top panel shows the axial magnetic field profile from a small NdFeB ring magnet (outer diameter 10.0~mm, inner diameter 6.7~mm, thickness 3.3~mm) placed a distance 7.4~mm from a 5~mm thick vapour cell. The magnet and vapour cell axial extent are shown by the blue and purple shaded regions, respectively. A non-uniform magnetic field (minimum 178~G, maximum 222~G) is produced across the cell by this configuration, which then yields the filter profile shown by the blue solid line on the bottom panel. The non-uniform field filter compares well to the filter profile used in previous work~\cite{Keaveney2016c} (dashed black line), with largely similar features and a slightly higher peak transmission value.}
	\label{fig:nonuniformfilter}
\end{figure}

In reference~\cite{Keaveney2016c}, a Faraday filter with high peak transmission was demonstrated with an approximately uniform axial magnetic field, produced by placing the 5~mm vapour cell between two NdFeB ring magnets, separated by a large distance compared to their extent, and the extent of the vapour cell. 
For applications development, using a smaller, single magnet system to generate the field is attractive for mechanical simplicity, miniaturisation purposes and cost-saving. However, using a single small magnet necessarily means that the field profile becomes non-uniform. However, as we show in figure~\ref{fig:nonuniformfilter}, we can effectively compensate for this field gradient with a suitable design of magnet, and ElecSus can be used as a design tool to optimise magnet specifications for this kind of application. In figure~\ref{fig:nonuniformfilter}, we simulate the filter profile that could be achieved with a small ring magnet placed close to the vapour cell. The magnet parameters were found by allowing the dimensions of the magnet (inner, outer diameter and thickness), the separation from the cell, and the cell temperature to vary, subject to some upper bounds on the magnet dimensions. For each iteration, the magnetic field profile over the cell and the resulting Faraday filter transmission calculated. The Faraday filter was then optimised for peak transmission at line centre using the methods outlined in refs.~\cite{Zentile2015f,Zentile2015d}. The simulation shows that a similar filter profile is generated from the non-uniform field (solid blue line in fig.~\ref{fig:nonuniformfilter}: field maximum/minimum over the cell: 222~G / 178~G), when compared with the filter profile used in ref.~\cite{Keaveney2016c} (dashed line in fig.~\ref{fig:nonuniformfilter}). Most of the features remain, and the peak transmission of the filter is slightly higher than the uniform field.

\section{Program structure}

\begin{figure}[t]
	\begin{center}
	\includegraphics[width=0.9\columnwidth]{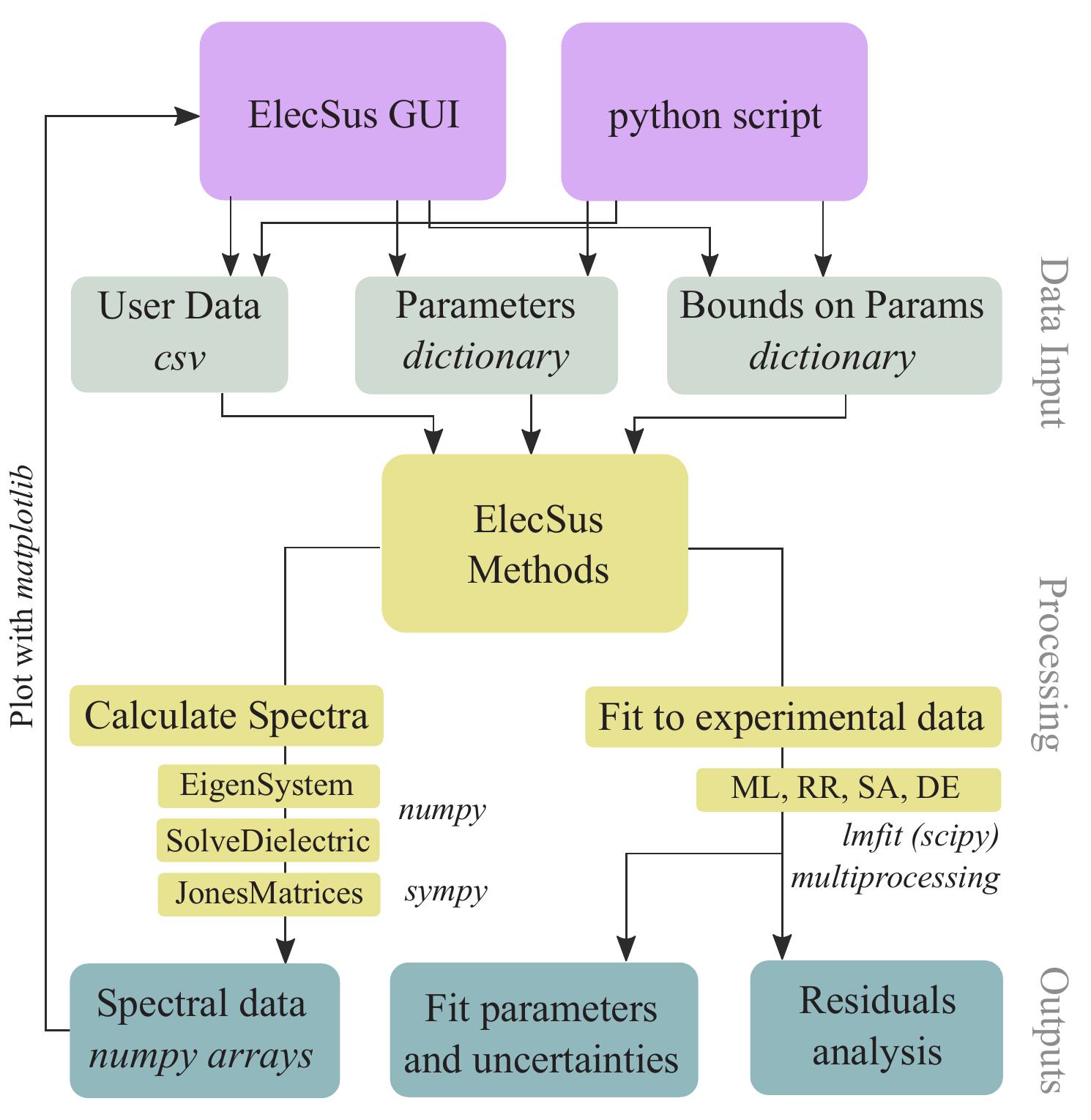}
	\end{center}
	\caption{Block diagram showing the flow of information in the ElecSus program. ML, RR, SA and DE refer to the fitting methods - see section~\ref{sec:fitting} for details.}
	\label{fig:FlowDiagram}
\end{figure}
The significant feature changes in ElecSus necessitated some changes to the overall program structure.
Figure~\ref{fig:FlowDiagram} shows a diagramatic illustration of information flow with the new program structure. 
The program can be accessed from either the GUI or an external Python script. In either case, the user supplies the simulation parameters, a set of exeperimental data if a fit is required and, optionally, the boundaries on fit parameters. 
These are passed to the {\tt calculate()} or {\tt fit\_data()} routines in {\tt elecsus\_methods.py}, which calculate spectra by finding the energy levels and state vectors of the system Hamiltonian ({\tt EigenSystem}), calculating the propagation of the electric field ({\tt SolveDielectric}) and finally applying the relevant Jones matrices to find the transmitted fields and intensities. The fitting methods have been updated and now use the {\it lmfit} module~\cite{Newville2014} which allows the use of boundaries for fit parameters - see below for more details.

\subsection{Graphical User Interface}

\begin{figure*}[t]
	\begin{center}
	\includegraphics[width=1.9\columnwidth]{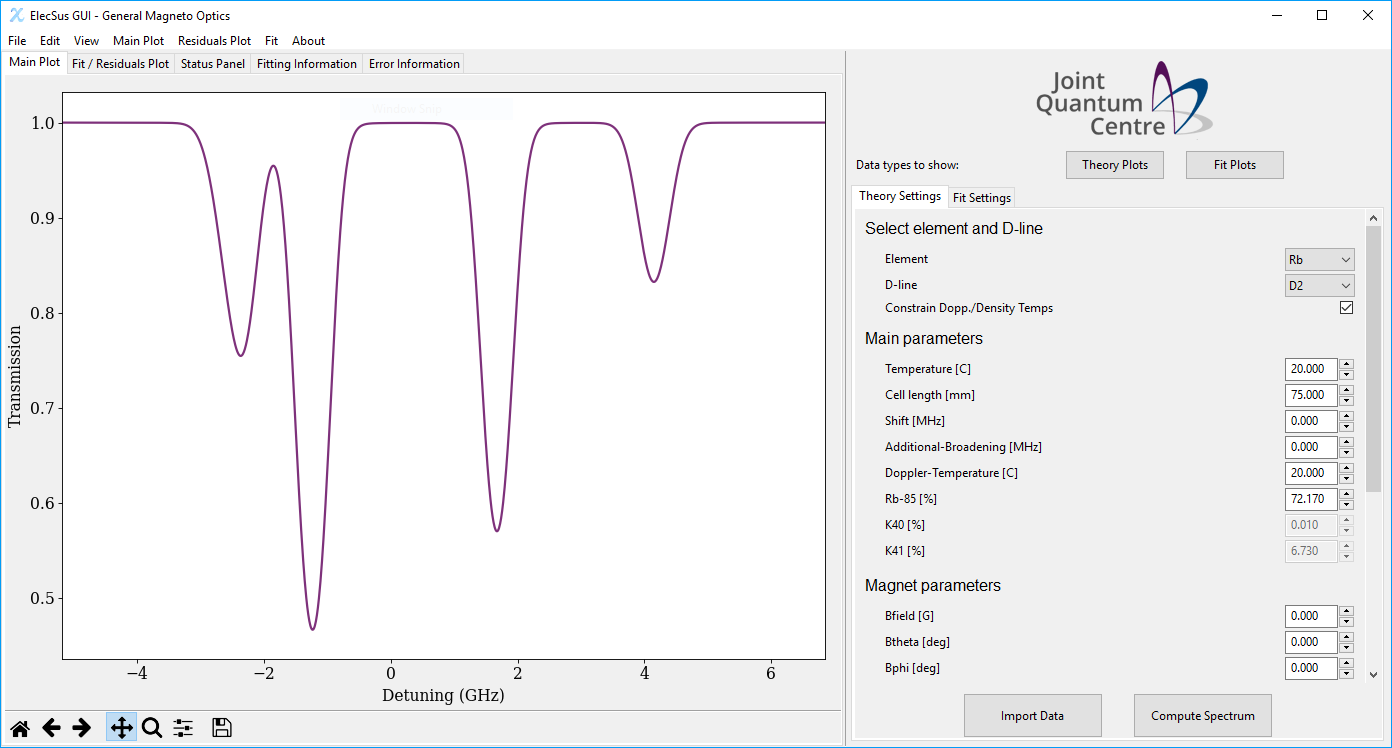}
	\end{center}
	\caption{Screenshot of the graphical user interface to ElecSus. The left side of the frame shows the calculated spectrum/spectra, whilst the right side contains the experimental parameters to simulate.}
	\label{fig:GUI}
\end{figure*}

\begin{figure}[t]
	\begin{center}
	\includegraphics[width=.95\columnwidth]{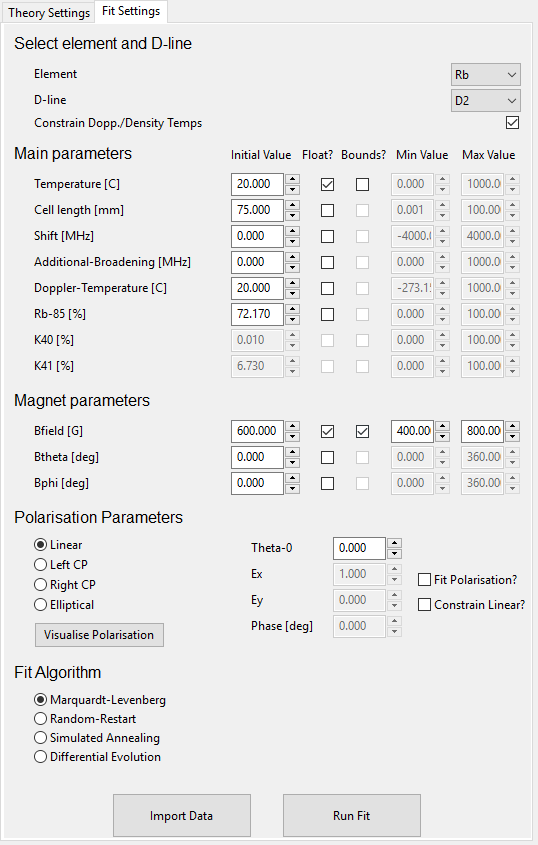}
	\end{center}
	\caption{Screenshot of the fitting options, showing the options to turn on/off fit parameters and add boundaries to fit parameters.}
	\label{fig:GUIfit}
\end{figure}

A graphical user interface (GUI) was developed for ElecSus which makes using the program much more accessible, particularly for users without knowledge of programming - it is now possible to use most of the program features without using any of the back-end source code. In figure~\ref{fig:GUI} we show a screenshot from the GUI, which is 
broadly split into two panels. On the left side, an interactive (i.e. axes can be dynamically rescaled) plot panel (using {\it matplotlib}~\cite{Hunter2007}) shows the spectral data that has been already calculated or loaded from user-supplied csv files. If fitting has been performed, a second tab in this panel shows the result of that fit, plotting residuals between the experimental and theoretical data. Additional tabs show text-based information about fit parameters, program status and any error messages that may have been generated.
On the right side of the window are the program input parameters, with two tabs for purely theoretical calculations or fitting data.
At the top of the panel, the user may select which outputs are displayed, from a list that includes all Stokes parameters, relative intensities of linearly polarised and circularly polarised light, and a few others. Underneath this are the input paramters. Figure~\ref{fig:GUI} shows the theory calculation panel, and figure~\ref{fig:GUIfit} shows the fit panel. For both, parameters are sub-divided into general parameters, and parameters specific to the magnetic field, and polarisation parameters.
Any fit parameter can be allowed to vary or be held constant, selected via the ``Float?'' tick-box. On selection, the fit bounds options become active, allowing the user either to avoid unphysical values, or to constrain some parameters to lie within some experimental uncertainty.
Finally, at the bottom of the panel, the user can select the fitting algorithm - see section~\ref{sec:fitting} for further details.

When experimental data is imported, it can be locally averaged (``binned'') or a moving-average smoothing applied using the Data Processing menu option. Data binning is recommended when the number of experimental data points is large ($\gtrsim 5000$), since the computation time scales roughly linearly with the data length.
After computing the spectra, the data can be exported, either by saving the plot as an image (in any of {\it matplotlib}'s supported formats: png, ps, eps, pdf, tiff, svg), or by saving the calculated data as a csv file.

\subsection{Methods file}

For integrating into other Python scripts, the {\tt elecsus\_methods.py} module
allows ElecSus to be called using a functional approach, for either calculation of spectra or fitting data using the {\tt calculate()} and {\tt fit\_data()} methods, respectively.

Experimental parameters are passed to these methods as {\tt key:value} pairs in a Python dictionary (a list of keys can be found in the code comments). This change has the advantage that parameters can be passed in any order, and unspecified parameters use default values, reducing the complexity of code needed.

\subsection{Update to the fitting methods}
\label{sec:fitting}

Though conceptually the same as the previous version, the implementation of data fitting has been updated for the new version, to make the code more clear and also to make use of new fitting options which are possible using the {\it lmfit} module~\cite{Newville2014}. This module natively supports the ability to fix or vary fit parameters, which greatly simplifies the coding required for a many-parameter fit where not all parameters are allowed to vary. {\it lmfit} also allows the user to specify bounds on parameters, which can prevent unphysical values (e.g. negative cell length) or narrow the parameter range when experimental details are known to a good level of accuracy.

The three algorithms from the previous version, Marquardt-Levenburg (ML), Random-Restart (RR) and Simulated Annealing (SA) are retained, and their functionality is the same. 
In addition to the above algorithms, Differential Evolution (DE) has been added as an option in the GUI, which is a global fitting routine based on stochastic methods developed by Storn and Price~\cite{Storn1997} that is reported to converge quicker than the SA (Metropolis algorithm~\cite{Metropolis1953}) method. In principle, any of the methods that are supported by the \textit{lmfit} module can be used. However, it is beyond the scope of this work to detail their individual advantages and disadvantages; more information can be found on the scipy documentation pages~\cite{ScipyMinimizeDocs}.
The ML method~\cite{Hughes2010} should be used for the simplest fit problems with few varying parameters. It is the quickest algorithm, but can only find local minima which is an issue for complex problems with a rich parameter space. In these cases, either DE, RR or SA should be used, which are more likely to find the global minimum of the parameter space.

\subsection{Additional changes since version 1}

Owing to the numerous additional features, the previous {\it runcard.py} way of using ElecSus is now unsupported, and hence backwards compatibility is broken with version 1. A full list of minor changes and updates can be found on the GitHub page for ElecSus.

\section{Installation and usage}
\label{sec:InstallAndUse}

The program is hosted on GitHub at
\begin{center} 
\url{www.github.com/jameskeaveney/ElecSus}
\end{center} 
and the program can be downloaded directly from there either by using {\tt git clone} if git is installed, or alternately as a zip archive. Installation as a Python module is optional, but can be done by running
\begin{center}
{\tt python setup.py install}
\end{center}
in a command-line/terminal from the top directory.
The GUI can be run from the command-line via
\begin{center}
{\tt python elecsus\_gui.py}
\end{center}
from the \texttt{elecsus} sub-directory.
For integration into user Python scripts, we provide {\tt elecsus\_methods.py} which includes two functions: {\tt calculate()} and {\tt fit\_data()}. These take in parameters as Python dictionaries (see source code doc-strings for lists of parameter keys), and output a series of numpy arrays which contain the spectral data, and fit parameters with associated uncertainties in the case of {\tt fit\_data()}.

\subsection{Test data}

Along with the program, we provide another GitHub repository which comprises a series of test data:
\begin{center}
\url{www.github.com/jameskeaveney/ElecSusTestData}
\end{center}
This test data includes the two examples from the previous version of ElecSus, and also includes all normalised experimental data from the figures in this paper. In the appendix we provide initial parameters for fitting ElecSus to these example data sets.

\section{Conclusions and outlook}
\label{sec:Conclusions}

We have presented an updated computer program to calculate the electric susceptibility of an alkali-metal vapour.
In addition to the previous features of ElecSus (versions 1 and 2), the program is now able to account for magnetic fields with arbitrary orientation with respect to the light propagation axis, and electric field propagation. Together, these allow calculation of susceptibility through non-uniform samples (e.g. magnetic field or density gradients), and the inclusion of optical elements such as birefringent windows. For each of these major changes, we have demonstrated their applications with comparison to example data sets, and find excellent agreement in all cases. In addition, we have developed a graphical interface and new API, which greatly simplifies the workflow for the majority of applications, which we hope will allow ElecSus to reach a wider audience and be more useful to the wider atomic physics community.

\section*{Acknowledgements}

We thank Nikola \u{S}ibali\'{c} and Daniel J. Whiting for helpful discussions and proof-reading the manuscript, Ilja Gerhardt for many stimulating discussions, Svenja Knappe for providing one of the compact vapour cells and Jacques Vigue for providing the 1.5~T magnet.
We acknowledge financial support from EPSRC (grant EP/L023024/1) and Durham University. In addition to the GitHub pages, the data presented in this paper are available from $<$DOI to be added at proof stage$>$.
%



\onecolumn
\appendix

\section{Jones matrices for common optical elements}

Table~\ref{tab:jones} lists Jones matrices for commonly used optical components in the $x,y$ basis. All elements are assumed to lie in the plane perpendicular to the propagation direction.

\begin{table*}[h]
	\begin{center}
		\begin{tabular}{Lc} 
\toprule
Optical component & Jones matrix \\
\midrule
\vspace{0.1cm}
Linear polariser aligned along $x$-axis 
			& \vspace{0.1cm} $\left(\begin{array}{cc} 1 & 0 \\ 0 & 0 \end{array}\right) $ \\

\vspace{0.1cm}
Linear polariser aligned along $y$-axis & 
			\vspace{0.1cm} $\left(\begin{array}{cc} 0 & 0 \\ 0 & 1 \end{array}\right) $ \\

\vspace{0.1cm}
Linear polariser aligned at angle $\theta$ to $x$-axis
			& \vspace{0.1cm} $\left(\begin{array}{cc} \cos^2(\theta) & \sin(\theta)\cos(\theta) \\ \sin(\theta)\cos(\theta) & \sin^2(\theta) \end{array}\right) $ \\

\vspace{0.1cm}
Left circular polariser
			& \vspace{0.1cm} $\frac{1}{2}\left(\begin{array}{cc} 1 & -i \\ i & 1 \end{array}\right) $ \\

\vspace{0.1cm}
Right circular polariser
			& \vspace{0.1cm} $\frac{1}{2}\left(\begin{array}{cc} 1 & i \\ -i & 1 \end{array}\right) $ \\


\specialrule{0.1pt}{0.2cm}{0.2cm}

\vspace{0.1cm}
Quarter-waveplate with fast axis at angle $\theta$ to $x$-axis
			& \vspace{0.1cm} $e^{i\pi/4} \left(\begin{array}{cc} \cos^2(\theta) + i\sin^2(\theta) & (1-i)\sin(\theta)\cos((\theta) \\ 
							(1-i)\sin(\theta)\cos(\theta) & \sin^2(\theta) + i\cos^2(\theta) \end{array}\right) $ \\

\vspace{0.1cm}
Half-waveplate with fast axis at angle $\theta$ to $x$-axis
			& \vspace{0.1cm} $ \left( \begin{array}{cc} \cos(2\theta) & \sin(2\theta) \\ \sin(2\theta) & -\cos(2\theta) \end{array}\right) $ \\

\specialrule{0.25pt}{0.2cm}{0.2cm}

\vspace{0.1cm}
Birefringent material that imprints a phase shift~$\phi$ oriented with the fast optical axis at an angle~$\theta$ to $x$-axis
			& 
			\vspace{0.1cm} 
			$ e^{-i\phi/2} \left( 
			\begin{array}{cc} 
				e^{i\phi/2}\cos^2(\theta) + e^{-i\phi/2} \sin^2(\theta) 
						& (e^{i\phi/2} - e^{-i\phi/2}) \cos(\theta) \sin(\theta) 
				\\ 
				 (e^{i\phi/2} - e^{-i\phi/2}) \cos(\theta) \sin(\theta) 
						& e^{i\phi/2}\sin^2(\theta) + e^{-i\phi/2}\cos^2(\theta)
			 \end{array}\right) $ \\
\bottomrule
		\end{tabular}
	\end{center}
	\caption{Jones matrices for common optical components in the $x,y$ basis.}
	\label{tab:jones}
\end{table*}

\newpage
\section{Test data information}

In table~\ref{tab:params} we list experimental parameters for the test data provided with the program. Suggested fit parameters are shown in bold for each data set.

\begin{table*}[h]
	\begin{center}
		\begin{tabular}{rccccccc} 
\toprule
Subfolder: & Faraday & Faraday & Faraday & Voigt
\\
File name (.csv): & S0\_RbD1 & S1\_RbD2 & Ix\_RbD2 & S0\_Voigt0,45,90
\\
Figure number: & Fig.~6 of \cite{Zentile2015b} & Fig.~7 of \cite{Zentile2015b} & Fig.~2 of \cite{Keaveney2016c} & Fig.~\ref{fig:VoigtS0} 
\\
\midrule
Data type: & $S_0$ & $S_1$ & $I_x$ & $S_0$
\\
Element: & Rb & Rb & Rb & Rb
\\
Isotopic abundance: & 1\% $^{85}$Rb & 1\% $^{85}$Rb & $^*$ & $^*$ 
\\
D-line: & D1 & D2 & D2 & D2 
\\
Polarisation: & Linear & Linear & Linear & Linear 
\\
$\theta_E$~(deg): & Any & {\bf 45} & 90 & 0,45,90 
\\
Cell length: (mm) & 1 & 1 & 5 & 1 
\\
$B$~(G): & {\bf 1000} & {\bf 5500} & {\bf 250} & {\bf 4200} 
\\
$\theta_B$~(deg): & 0 & 0 & 0 & 90  
\\
$\phi_B$~(deg): & 0 & 0 & 0 & 0 
\\
Temperature ($^\circ$C): & {\bf 130} & {\bf 65} & {\bf 92} & {\bf 100} 
\\
Additional-Broadening (MHz): & {\bf 5} & {\bf 5} & 0 & 0 
\\
Suggested fit method: & ML & RR & ML & DE 
\\
\midrule
\\
Subfolder: & Birefringence & Birefringence & Arbitrary Angle & Field Gradient
\\
File name (.csv): & S1\_Voigt\_Biref & S3\_Voigt\_BirefCorrected & S0\_Btheta120.csv & S0\_Bgradient
\\
Figure number: & Fig.~\ref{fig:VoigtS1} & Fig.~\ref{fig:VoigtS3} & Fig.~\ref{fig:arbangle} & Fig.~\ref{fig:gradient}
\\
\midrule
Data type: & $S_1$  & $S_3$ & $S_0$ & $S_0$
\\
Element: & Rb  & Rb & Rb & Rb
\\
Isotopic abundance: & 1\% $^{85}$Rb & 1\% $^{85}$Rb & $^*$ & $^*$
\\
D-line:  & D2 & D2 & D2 & D2
\\
Polarisation: & Linear & Linear & Linear & Linear
\\
$\theta_E$~(deg): & 45  &  45 & {\bf 8} & 45
\\
Cell length (mm): & 1 & 1 & 1 & 75
\\
$B$~(G): & {\bf 15400}  & {\bf 15400} & {\bf 4200} & { 1000}$^\dagger$
\\
$\theta_B$~(deg): & 90 & 90 & {\bf 120} & 0
\\
$\phi_B$~(deg): & 0 & 0 & 0 & 0 
\\
Temperature ($^\circ$C): & {\bf 98} & {\bf 125} & {\bf 90} & {17.5}
\\
Additional-Broadening (MHz): & 40 & 40 & 15 & 15
\\
Suggested fit method: & DE & DE & DE & ML
\\
\midrule
\bottomrule
		\end{tabular}
	\end{center}

	\caption{Parameters for supplied sample data. Bold indicates suggested parameters to vary for fitting. $^*$: natural abundance. $^\dagger$: field is non-uniform; this is the average field.}
	\label{tab:params}
\end{table*}

\twocolumn
\section*{References}

\bibliographystyle{elsarticle-num}
\bibliography{library}







\end{document}